  \documentstyle[12pt]{article}
\begin{document}
  \baselineskip=24pt
  \parskip=0pt plus2pt
  \textheight=22cm

 \begin{titlepage}
 \setcounter{page}{0}
 \begin{center}
 {\LARGE\bf Model for Glass Transition in a Binary fluid from a Mode
 Coupling approach}
 \end{center}
 \vspace{.5in}
 \begin{center}
 {\it Upendra Harbola and Shankar P. Das}\\
 {\it School of Physical Sciences,\\
 Jawaharlal Nehru University, \\
 New Delhi 110067, India}

 \vspace*{2cm}
 {\large ABSTRACT}
\end{center}
 \noindent
 We consider the Mode Coupling Theory (MCT) of Glass
  transition for a Binary
  fluid. The Equations of Nonlinear Fluctuating Hydrodynamics
  are obtained with a proper choice of the slow variables
  corresponding to the conservation laws. The resulting
  model equations are solved in the long time limit
  to locate the dynamic transition. The transition point
 from our model is considerably higher than predicted
 in existing MCT models for binary systems. This is in
  agreement with what is seen in Computer Simulation of binary
 fluids.

\noindent

\vspace*{5cm}
\noindent

PACS number(s) : 64.70P, 05.60.+w, 64.60C, 47.35.+i,05.20.-y

\end{titlepage}

The phenomenon of glass transition has been studied widely in recent years
using the mode coupling models for dense liquids.
This involves a feed back to the transport properties of the liquid
from the coupling of hydrodynamic  modes in the liquid. 
The key aspect of the Mode Coupling Theory \cite{gene} (MCT) of 
Glass transition takes into account the enhanced feed back
effects to the viscosity due to the coupling of the slowly decaying density
fluctuations. 
As the density increases beyond a critical value the system undergoes 
a dynamic transition to a non-ergodic state. The long time limit of the 
 time correlation of density fluctuations is treated as an
order parameter for structural relaxation. 
The transition to the nonergodic state is characterized by a nonzero
value of the order parameter or the so-called Non Ergodicity Parameter
(NEP).
For a binary mixture the self consistent MCT for Glass Transition 
and the dynamic instability resulting from the feedback mechanism has 
been studied by several authors \cite{TKBT}-\cite{FL} in the past.
The model equations in these works predict the dynamic transition at
a much lower density compared to the one where a structural arrest is
seen in computer simulation of similar systems.
This aspect of the Mode coupling model for the binary fluid
was indicated in the computer simulations results
reported in Ref. \cite{WK}.
In the present work we have constructed the MCT for the binary
system from the equations of the
Nonlinear Fluctuating  Hydrodynamics(NFH) \cite{ted,DM} with
proper conservation laws.
We consider the correction
to the transport coefficient using the mode coupling approximation
of dominant density fluctuations. This gives rise to the feed back
mechanism on the transport properties.
The possible dynamic transition allowed by the model equations
is analyzed in terms of the solution of the resulting integral
equations.
Our main findings are very different from the existing 
 results \cite{TB}. 
 The theoretical prediction for the critical density of the
 dynamic transition now shifts to much higher density
 bringing it in better quantitative
agreement with the computer simulation results.

 The slow variables for a binary fluid are partial densities 
 $\rho_s(\vec x)$, $s=1,2$ and total momentum density $\vec g (\vec x)$.
 These can be defined microscopically as,
 \begin{eqnarray}
 \label{varb}
 \rho_s (\vec{x})&=& m_s
 \sum_{\alpha=1}^{N_s} ~ \delta(\vec{x}-\vec{R_{\alpha}^s}(t))
 \nonumber \\
 \vec g_i (\vec x)&=&\sum_{s=1}^{2}
 \sum_{\alpha=1}^{N_s} \vec P_{i\alpha}^s(t)~
 \delta(\vec{x}-\vec{R_{\alpha}^s}(t))
 \end{eqnarray}

 \noindent
 where $\vec R_{\alpha}^s(t)$ and $\vec P_{i\alpha}^s(t)$
 are respectively the position and $i$-th component of the momentum of the
 $\alpha$-th
 particle in the $s$-th species.
 $N_s$ is the number of particles in the $s$-th species in the
 mixture and the total number of particles is given by,
 $ N= \sum_s N_s $.
 \noindent
 Generalized Langevin equation for the slow variables are
 \begin{equation}
 \label{deq}
 \frac{\partial \rho_{s}}{\partial t}+ \nabla .[\frac{\rho_s}{\rho} \vec g]
 + ~\gamma_{ss\prime} \frac {\delta F_u}{\delta \rho_{s\prime}}= \theta_s
 ~~~~~~~~~ s=1,2
 \end{equation}
 \begin{equation}
 \label{momeq}
 \frac{\partial g_i}{\partial t}+\nabla_j \frac{g_i g_j}{\rho}
 +\rho_s \nabla_i \frac{\delta F_u}{\delta \rho_s}+
 L_{ij} \frac{\delta F}{\delta g_j} = f_i
 \end{equation}
 \noindent
 where the repeated indices are summed over - a convention that we
 follow in the rest of the paper, unless specifically mentioned.
 The $\gamma_{s\prime{s}}$ correspond to the inter and
  Self-diffusion in the two species.
 We use the expression \cite{supurna} $ \gamma_{s\prime{s}}
 \equiv {\partial}_i \gamma_{s\prime{s}}^{ij} {\partial}_j $.
 On adding the equations (\ref{deq}) with $s=1,2$ for the two components,
 the continuity equation  
${\partial \rho} / {\partial t}+ \nabla .\vec{g}=0$
for the total density
 $\rho $ ( = $ \rho_1 + \rho_2 $ ) is obtained. 
Thus the diffusion kernel in the first two equations of (\ref{deq})
 for $s=1,2$ should cancel. We will use the simple choice here that
 $\gamma_{21} = - \gamma_{11} \equiv \gamma_o  \nabla^2 $ and
 $ \gamma_{12} = - \gamma_{22} \equiv \gamma_o  \nabla^2 $.
 Similarly the noise corresponding to density fluctuation satisfies
 $\theta_1+\theta_2 =0 $.
 Noise is assumed to be white Gaussian and related to the corresponding bare
transport coefficients. The $f$ and $\theta$ are taken to be
uncorrelated. 
 The bare transport coefficient $L_{ij}$ can be expressed in terms 
 of the shear and bulk viscosities $\eta_o$ and $\zeta_o$ 
  in an isotropic fluid. 
 The longitudinal viscosity is given by $\Gamma_o=(4\eta_o/3+
 \zeta_o)/\rho_o$.

 The Free energy functional $F$ that is used in constructing the
 equations of the Nonlinear Fluctuating Hydrodynamics plays an important
 role here. It has two parts, the
 kinetic and the "potential" respectively denoted by $F_k$ and $F_u$.
 In computing the kinetic energy  contribution, we follow
 the standard procedure due to Langer and Turski \cite{LT}.
 \begin{equation}
 \label{kinetic}
 F_k=\frac{1}{2} \int d \vec x [ {g^2 (\vec x)} / {\rho(\vec x)} ]
 \end{equation}
 \noindent
 where the total density $\rho = \rho_1 + \rho_2 $ appears in the
 denominator of (\ref{kinetic}).
 For the potential part of the free energy there is
 an ideal gas contribution together with the interaction term.
 We use the standard form \cite{rama-yu,supurna} 
in terms of the direct correlation functions \cite{rama-yu},
 \begin{equation}
 \label{potential}
 F_u( \rho) = \frac{1}{m_s} \int d \vec x \rho_s(\vec{x})
 ~ [{\rm ln} \frac{ \rho_s(\vec x)} {\rho_{os}}-1]
 - \frac{1}{2 m_s m_{s\prime}} \int d \vec x d \vec x \prime
 c_{ss \prime} (\vec x-\vec x \prime) \delta
 \rho_s(\vec x) \delta \rho_{s \prime} (\vec x \prime)
 \end{equation}

 \noindent
 From the dynamic equations (\ref{deq}) and (\ref{momeq}),
 we compute the time correlation function \cite{MRT} 
$C_{ss\prime}(\vec x-\vec x\prime,t - t^{\prime})=
 \frac{1}{N}<\psi_s(\vec x,t) 
\psi_{s\prime}(\vec x\prime, t\prime)>$ 
between the hydrodynamic variables $\psi_s$ and $ \psi_{s\prime}$. 
 The normalized form of the
 density-density correlation functions, $R_{s s\prime}(q,z)$, is
defined as $ R_{s s\prime}(q,z)= 
C_{s s\prime}(q,z)/ \sqrt{\chi_{ss}\chi_{s\prime s\prime}} $
 where $\chi_{s s\prime}$ is the equal time density correlation function
 between i-th and j-th species.
 The partial structure factor $S_{ss'}$ \cite{AL} is defined from the
 $\chi_{ss'}$
 through the relation $\chi_{ss'} = a_{s} a_{s\prime} S_{ss\prime}$
 where  $a_s = m_s\sqrt{n_s}$ ,
 $m_s$ and $n_s$ are respectively mass and particle density corresponding to
 the s-th species.
 We compute the mode coupling contribution to the transport coefficients 
 by treating the non linearities in equations of motion with standard 
 methods \cite{MRT}.
 The renormalized form for the longitudinal viscosity at the one loop
 order is given by,
 \begin{equation}
 \label{gamma}
 \Gamma (q, t)= \frac{1}{2\rho_0 q^2 }
 \int \frac{d\vec{k}}{(2\pi)^3}
 V_{ss \prime}(q,k)~V_{ll\prime}(q,k_1)
 ~C_{l\prime s\prime}(k, t)~C_{ls}(k_1, t),
 \end{equation}

 \noindent
 where $k_1=q-k$. The vertex function $V_{ss\prime}(q,k)$ is given by,
 \begin{equation}
 \label{vertex}
 [m_s m_{s\prime}]
 V_{ss\prime}(q,k)= 
 (\widehat q .k) ~~{\tilde{c}}_{ss\prime} (k) + \widehat q .k_1~~
 {\tilde{c}}_{ss\prime}(k_1) 
 \end{equation}

 \noindent
with  $ c_{ss \prime}(\vec x-\vec x \prime)$  denoting
 the equilibrium two particle
 direct correlation function \cite{hansen}.
 The quantity $ \tilde{c}_{ss\prime}(q) =
 (n_s n_{s\prime})^{\frac{1}{2}}
 c_{ss\prime}(q) $ is related to the structure factor
 $S_{ss\prime}(q)$ through the Ornstein Zernike relation
 $ [\delta_{ls} - \tilde{c}_{ls} (q) ] S_{ls\prime}(q)
 =\delta_{ss\prime}$.
 In obtaining the above expression(\ref{gamma}) we have considered 
 only the coupling of the density fluctuations coming from the non 
 linearity in the Pressure term that appears in the momentum conservation
 equation (\ref{momeq}).
 This involve taking the density fluctuations as dominant and using the one
 loop correction or the so-called Kawasaki approximation
 to the four point functions to the transport coefficients.
  This is done in the same spirit of the self-consistent mode coupling
  approximation of taking density fluctuations to be dominant as in the
  case of one component fluid.
 We like to point out here that in the appropriate limit
 these results reduce to the one component fluid result that
 {\em is in complete agreement} with all other wave vector dependent
 models \cite{beng} for one component systems.

 Next we consider the implications on the  dynamic transition
 as a result of the new set of integral equations for the NEP.
 The ideal glass phase is characterized by  the nonergodicity parameter,
 $\lim_{t\rightarrow \infty} R_{s s\prime}(q,t)= f_{s s\prime}(q)$.
 The correlation of density fluctuations is expressed in terms of the
 renormalized transport coefficients \cite{DM}.
 In the asymptotic limit of long times 
  we obtain the following
 set of self-consistent equations for the NEP's $f_{s s\prime}(q)$,
 \begin{equation}
 \label{fqq}
 f_{ss\prime}(q)= \frac{\Im_{ss\prime}(q) \Gamma(q)}
 {1+\Omega(q)\Gamma(q)} \ \ \ .
 \end{equation}

 \noindent
 Here $ \Gamma(q) $ is the long time limit of
 $ \Gamma(q, t) $ given by equation (\ref{gamma}). $\Omega(q)$
 and $\Im(q)$ are given by,
 \begin{equation}
 \label{renom}
 \Omega(q) = \Delta_{ss\prime} S_{ss\prime}(q)  \ \ \ \ \ \ \ \
 \Im_{ss\prime}(q)
 = \frac{\Delta_{ij} S_{is}(q) S_{j s\prime}(q)}
 {[S_{ss}(q)S_{s\prime s\prime}(q)]^{\frac{1}{2}}}
 \end{equation}

 \noindent
 with $\Delta_{s s\prime} = a_s a_{s\prime} / \rho_o $.
Eq. (\ref{fqq}) constitute a set of coupled nonlinear integral
 equations for the non-ergodicity parameters $f_{s s\prime}$.
 The dynamic instability of the ideal glass transition in the
 binary system is then located from the {\em self-consistent}
 solution of eq. (\ref{fqq}) by  iterative method in a similar
  manner as in the one component systems.
 The static structure factor goes as an input in the calculation.
 The interaction potential for the system enters the present 
 theoretical description for the dynamics through this thermodynamic
 quantity.
 Due to the available simple formulation for the structure and in order to
 keep continuity with pervious works on binary fluids \cite{TB}
 we will choose a mixture of hard spheres. This is described by
 three independent parameters,
(a)
 The fractional concentration of particles $ x_s (=N_s/N)$, 
(b) The size ratio $\alpha(=\sigma_1/\sigma_2)$,
and 
(c)
 The total packing fraction $\eta=\eta_1+\eta_2$ where
 $\eta_1$ and $\eta_2$ are the packing fractions of the individual
 species. $\eta_s=\frac{\pi}{6} n_s{\sigma_s}^3$,  $n_s$ and
 $\sigma_s$ being respectively the number of particles per unit
  volume and diameter of the s-th species  (s =1,2). We choose
 $\sigma_2$, diameter of bigger species, as the unit of length.
 The structure factors for the binary liquid required
 in computing the mode coupling vertex functions
   appearing in (\ref{fqq}) are obtained using the
 standard results of the solution of
 Percus Yevick equations for the hard sphere \cite{lebowitz,AL} mixture.
 We solve for the nonlinear integral equations to search the
  nontrivial fixed points with non zero values of
 $ f_{ss\prime}(q)$. Depending on the thermodynamic parameters
   described above a nonergodic phase is seen beyond a critical density.
 The results remain unchanged if we replace $x$ and $\alpha$
  respectively by $1-x$ and $1/\alpha$ meaning an
  interchange of the label for the two species. In Fig. 1,
 we show the non-zero solutions for the NEPs $f_{11}$ and
  $f_{22}$ respectively for the critical packing fraction $\eta_c =.59$. 
 The nature of the ideal glass instability in the binary mixture is 
dependent on the  size ratio $\alpha$ as well as the relative abundance 
$x$. 
We list below  a few comments.

\begin{itemize}
\item[(i)]
The NEP $f_{22}$ reach a small value at an intermediate wave number $q_m$ 
 depending on $x$ (fraction of the bigger particles 2). 
As the relative proportion $x$
is increased this wave number $q_m$ become 0, beyond
 a value $x\geq x_0$. This behavior is primarily due to the
 nature of the structure function $\Im_{ss\prime}(q)$ that appears in
 the mode coupling equation (\ref{fqq}) for the NEPs.
 In the inset of Fig. 1, we show how the position of $q^*_m$
 shifts with x at $\alpha=.7$ and becomes zero at $x_0=.977$.

 \item[(ii)]
 The transition to the non-ergodic phase, indicating a
 structural arrest, shifts to higher densities as
 the size ratio $(\alpha)$ is decreased. For example, at $x=.1$
 and $\alpha=.2$
 {\em structural arrests are absent} up to very high densities
 ($\eta \leq .64$). 
This absence of transition here can be explained from the fact that
for low values of $x$ and size ratio $\alpha$ the smaller
 particles can diffuse easily through the cage of the bigger particles
 and hence the structural arrest is avoided.
 Indeed it is a competition between the two quantities, the size ratio
 $\alpha$ allowing easier movement of the particles and
 $x$ giving the relative abundance of the voids formed between
 bigger particles.
To stress the importance of the present work, it should
 be noted that, with the same values of the thermodynamic
 parameters the Ref. \cite{TB} report a transition in the range
 ($.51 \leq \eta_c \leq .52$).

 \item[(iii)]
  With the ratio of particle sizes, $\alpha$ smaller than .75, there
 is no structural arrest, over a range $(.22 \leq x \leq .58)$.
 In the earlier mode coupling model equations \cite{TB} this sensitivity to
 the sizes of the particles ($\alpha$) and ($x$) was absent.
 The same qualitative dependence of $\alpha$ on the freezing
 in a binary fluid system has also been reported in 
 studies related to the thermodynamics \cite{BBH} of the system.
 \end{itemize}

\noindent
 Fig. 2 shows the critical packing fraction $(\eta_c)$ with x at constant
 $\alpha=0.8$.
 If the particles are similar to each other $(\alpha=1)$ and x close to unity, 
 the dynamic instability occurs at $\eta=0.515$ 
that is very close to the critical packing
 fraction ( with the use of Percus-Yevick Structure factor )
 for a single component system as is expected to be the case.

 As a result of the mode coupling instability, the transverse current
 correlation function \cite{supurna}
 also develops propagating shear modes \cite{rajeev}. The expression
 for the speed of the shear waves ($c_T$) in the
 binary fluid can be obtained in terms of  nonergodicity parameters
 in the following form,
\begin{equation}
\label{sher-sp}
{c_T}^2= \frac{1}{60\pi^2}\sqrt{\frac{x_1x_2}{n_1n_2}} \int dk k^4
\tilde{c\prime}_{ll\prime}(k)
\tilde{c\prime}_{ss\prime}(k) f_{s\prime l\prime}(k)f_{sl}(k)
\sqrt{S_{ss}(k)S_{ll}(k)S_{s\prime s\prime}(k)S_{l\prime l\prime}(k)}
\end{equation}

\noindent
where $\tilde{c\prime}_{ss\prime}(q)$ denotes derivative of
$ \tilde{c}_{ss\prime}(q)$ with respect to q.
In Fig. 3, we show the behavior of $c_T$ with $(x)$ for two different 
size ratios($\alpha$).

 In the present work we have obtained, from the equations of Nonlinear
 fluctuating hydrodynamics, the feedback mechanism of the
  self consistent MCT. Solutions of our model equations indicate that
 the dynamic instability {\em shifts to much higher
 densities} than what is predicted in the earlier works
 \cite{TB}.
This also agrees with the trend seen in
 computer simulation results\cite{WK}.
The kinetic energy term, eq. (\ref{kinetic}),
 of the Free energy functional that is used here
 is crucial for the construction of the model equations.
 This is also essential for obtaining in the momentum
   conservation equation the convective
 or the so-called Navier-Stokes term reflecting Galilean invariance.
 In considering the dynamic nonlinearities
 that gives rise to the mode coupling instability
 we have taken here a driving Free energy functional of
 Gaussian form. Thus  origin of the nonlinearity is dynamic as
 is the case for the one component fluid.
 The higher order terms in the Free energy functional will give rise
 to nonlinearities in the density equations and this will require for the
 renormalization  of the diffusion coefficients ($\gamma_o$).
 The lowest order vertex then is proportional to the $\gamma_o$ and
 will come from non-Gaussian terms in the free energy functional involving
 higher order direct correlation functions.
 We end the paper with few comments,
 \begin{itemize}
 \item[(i)]
 The earlier versions of
  the MCT \cite{TB} for binary systems can be reproduced from our work,
 using the {\it same Gaussian Free Energy functional} $F_u$as considered here,
  if one (a)  replaces the  kinetic energy term  with \cite{supurna}
 $ F_k= \int d \vec x [ g_1^{2}/2\rho_1 + g_2^{2}/2\rho_2 ] $,
 $\vec{g_1}$ and $\vec{g_2}$ denoting the momentum densities of the
 individual components in the binary system, both treated as two 
 {\em separately} conserved variables and
 (b) ignore the inter diffusion $\gamma$ as well. (a) will also
 violate the Galilean invariance in the hydrodynamic equations.
 
 \item[(ii)]
Here we have considered the dynamics of density correlation functions
only in the asymptotic limit so as to study the implications on the
dynamic transition.
 The utility of the present approach is demonstrated through
 the better result for the location of the dynamic transition.
 To study the dynamics over different time scales the time dependent
 mode coupling equations for the various density correlation functions
 has to be considered. These equations involve the bare 
 transport coefficients $\gamma_{ij}$ and $L$ and will be reported
 elsewhere \cite{up-spd2}.  

  \item[(iii)]
 Finally, the complete picture
  of the dynamics will involve computing the coupling to the
  current correlations \cite{DM} that restores the ergodicity  in the
  longest time scale \cite{up-spd2}.
 The present work treats the dynamics for the two
 component system with the correct set of slow variables
 and shows quantitative agreement of
 MCT results with computer simulation studies and remedies the theories
 presented in earlier works.
 \end{itemize}

 \vspace*{.5cm}
 \noindent
 \section*{Acknowledgement}
SPD acknowledges the support from grant INT9615212 from NSF. UH
acknowledges financial support from University Grant Commission,
India.

 \vspace*{.5cm}
 \noindent
 \section*{Figure Captions}

 \vspace*{.5cm}
 Fig. 1 :
 nonergodicity parameters
 $ f_{11}(q^*) $(solid line) and $ f_{22}(q^*)$
 (broken line) at $\eta_c=.59$, x=.8 and $\alpha=.7$.
 $q^*$=q$\sigma_2$ is plotted along x axis.
 Inset shows $q^*_m$(see text) Vs x plot for $\alpha=.8$
and $\eta=.59$

 Fig. 2 : critical value of $\eta_c$ is plotted
with $x$ for $\alpha=0.8$. Regions "a" and "b" represent
liquid and glass phases respectively.

 Fig. 3 : speed of shear mode $c_T$ is
 plotted as a function of x for two values
  of $\alpha$=0.85 (solid) and 0.90 (broken)  at $\eta=.57$.

 
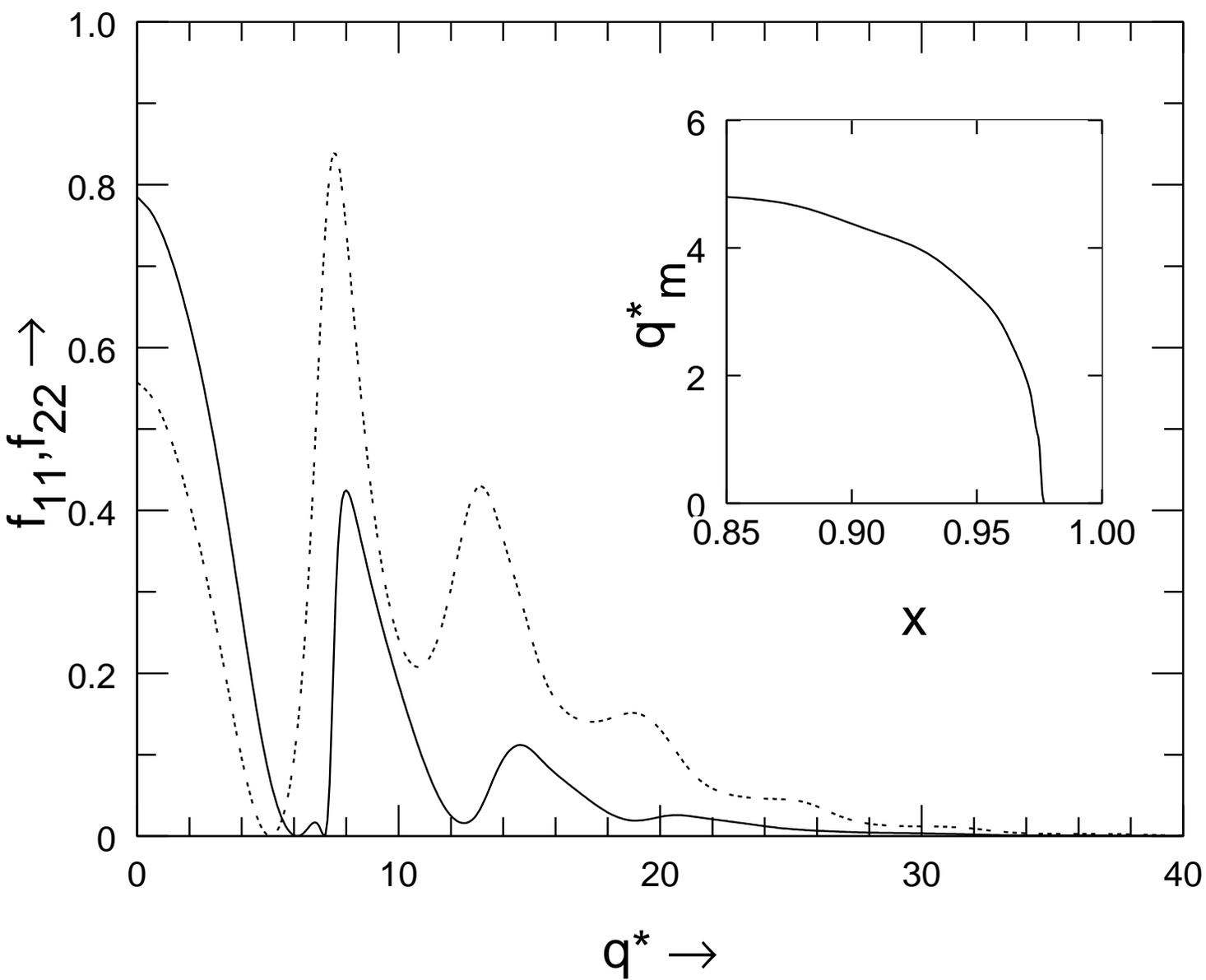
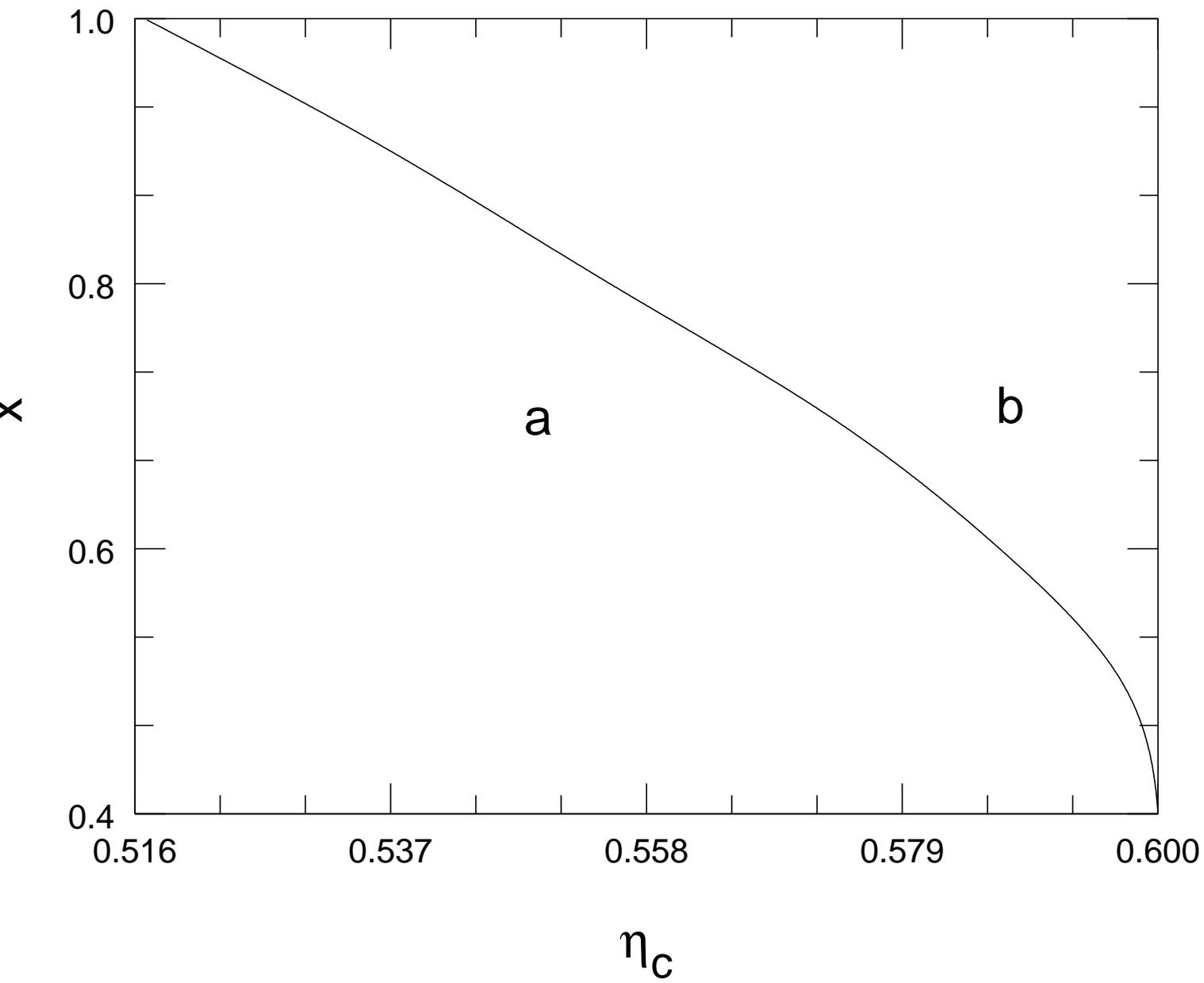
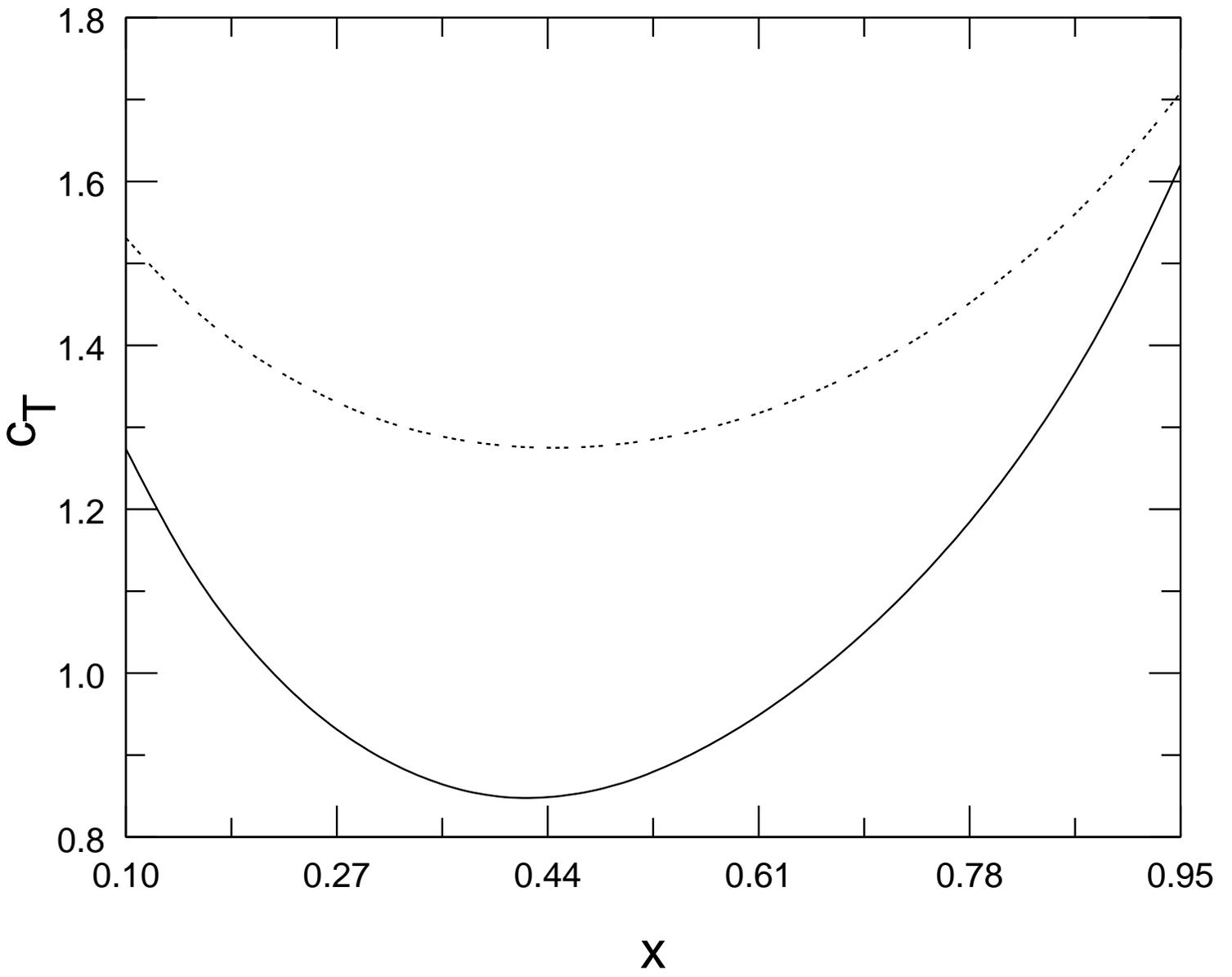
\end{document}